\documentclass[letter,comsoc,10pt,twocolumn]{IEEEtran}
\usepackage{latexsym}
\usepackage{amsfonts}
\usepackage{amssymb}
\usepackage{amsbsy}
\usepackage{mathtools}
\usepackage{amsthm}
\usepackage{color}
\usepackage{tikz}
\usetikzlibrary{automata,arrows,positioning,calc}
\usepackage{pgf}
\usepackage{graphicx,lscape,rotating,subfigure}
\usepackage{array}
\newcolumntype{L}[1]{>{\raggedright\let\newline\\\arraybackslash\hspace{0pt}}m{#1}}
\newcolumntype{C}[1]{>{\centering\let\newline\\\arraybackslash\hspace{0pt}}m{#1}}
\newcolumntype{R}[1]{>{\raggedleft\let\newline\\\arraybackslash\hspace{0pt}}m{#1}}

\usepackage{balance}
\usepackage[varg]{txfonts}
\usepackage{algorithm,algpseudocode}
\usepackage[noadjust]{cite}
\usepackage{multirow}

\theoremstyle{remark}

\theoremstyle{definition}

\title{\Large Analysis of Latency and MAC-layer Performance for Class A LoRaWAN}

\author{\IEEEauthorblockN{Ren\'{e} Brandborg S{\o}rensen,~\IEEEmembership{Student Member, ̃IEEE,} Dong~Min~Kim,~\IEEEmembership{Member, ̃IEEE,} Jimmy~Jessen~Nielsen,~\IEEEmembership{Member, ̃IEEE,} Petar~Popovski,~\IEEEmembership{Fellow, ̃IEEE}}%
\thanks{This work has been supported by the European Research Council (ERC Consolidator Grant Nr. 648382 WILLOW) within the Horizon 2020 Program.}
\thanks{All authors are with the Department of Electronic Systems, Aalborg University, Denmark (Email: \{rbs,dmk,jjn,petarp\}@es.aau.dk).}%
}

\begin{document}
\maketitle

\begin{abstract}
We propose analytical models that allow to investigate the performance of Long Range
Wide Area Network (LoRaWAN) uplink in terms of latency, collision rate, and throughput under
the constraints of the regulatory duty cycling, when assuming exponential inter-arrival
times. Our models take into account sub-band selection and the case of sub-band
combining. Our numerical evaluations consider specifically the European ISM band, but the analysis is applicable to any coherent band. Protocol simulations are used to validate the proposed models. We find that sub-band
selection and combining have a large effect on the QoS experienced in a LoRaWAN cell
for a given load. The proposed models allow for optimizing resource allocation within a
cell given a set of QoS requirements and a traffic model.
\end{abstract}

\begin{IEEEkeywords}
LoRa; LoRaWAN; LPWA; IoT; QoS; latency; duty cycle; low power; long range.
\end{IEEEkeywords}



\section{Introduction}

Services utilizing communications between machines are expected to receive a lot of
attention, such as health monitoring, security monitoring and smart grid services
\cite{palattella2016internet}. These Internet of Things (IoT) services generate new
demands for wireless networks. The spectrum of service scenarios in the IoT is wide and
as a result the required quality of service (QoS) across IoT services is also wide. In
some scenarios ultra high reliability is required, in others a low latency is required
and supporting massive numbers of low-cost and low-complexity devices is still important
issue. The devices can be served by the cellular networks and, specifically, by their
M2M-evolved versions, such as Narrowband IoT (NB-IoT) \cite{wang2016primer}. However,
there is a low-cost alternative for serving these devices using Low Power Wide Area
(LPWA) networks that operate in unlicensed bands. The number of IoT devices connected by
non-cellular technologies is expected to grow by 10 billions from 2015 to 2021
\cite{ericsson2016mobile}. It is therefore of interest to develop QoS models for the LPWA
protocols in order to analyze which protocol is best suited for a given service.

Long Range Wide-area Network (LoRaWAN) is an emerging protocol for low-complexity
wireless communication in the unlicensed spectrum using Long Range (LoRa) modulation.
The scalability and capacity of LoRaWAN is investigated in
\cite{mikhaylov2016analysis} where it is implicitly assumed that the inter-arrival times are fixed.
In \cite{scaleBor} the scalability is evaluated in terms of goodput and network energy consumption. One of the key elements of LoRaWAN is
the use of duty cycling in order to comply with the requirements for unlicensed
operation. Duty cycling is imposed per sub-band by regulation and optionally also aggregated for all
bands. It is the central factor that sets limitation on the throughput and the latency of
the network.
%
The limits of duty-cycled LoRaWAN are pointed out in \cite{adelantado2016understanding}, but only aggregated duty cycle and fixed inter-arrival arrivals are considered.

The contribution of this paper is an analytical model of the LoRaWAN uplink (UL) that characterizes the performance, in terms of latency and collision rate,
under the influence of regulatory and aggregated duty cycling, assuming exponential inter-arrival times.
The obtained latency and collision rate results from the analysis are verified through simulation.

We summarize the key features of LoRaWAN in Section~\ref{sec:lora}. A system model is
presented in Section~\ref{sec:model} and analysed in Section~\ref{sec:ana}. Numerical
results based on the analysis and simulation is shown in Section~\ref{sec:eval}.
Concluding remarks are given in Section~\ref{sec:conc}.

\section{Long Range Wide Area Network} \label{sec:lora}

LoRaWAN is a wireless communication protocol providing long range connectivity at a low
bit rate. LoRaWAN is based on the LoRa modulation.
LoRaWAN supports LoRa
spreading factors 7 to 12. The overhead of a LoRaWAN message with a payload and no
optional MAC command included is 13~bytes.

LoRaWAN defines a MAC layer protocol to
enable low power wide area networks (LPWAN) \cite{loraspec20015}. A gateway serves
multiple devices in a star topology and relays messages to a central server. LoRaWAN
implements an adaptive data rate (ADR) scheme, which allows a network server to select both the data rate and the channels to be used by each node.

Three different classes (A, B and C) of nodes are defined in LoRaWAN. Class A has the
lowest complexity and energy usage. All LoRaWAN devices must implement the class A
capability. A class A device can receive downlink messages only in a receive window.
There are two receive windows after a transmission in the uplink. The first window is
scheduled to open 1 to 15~second(s) after the end of an uplink transmission with a negligible 20~ms margin of error. The second window opens 1~second after the end of the first.

LoRaWAN utilizes the industrial, scientific and medical (ISM) radio bands, which are
unlicensed and subject to regulations in terms of maximum transmit power, duty cycle and
bandwidth. 
The end-device also obeys a duty cycling mechanism called the aggregated duty
cycle, which limits the radio emission of the device. An aggregated duty cycle of 100~\%
corresponds to the device being allowed to transmit at any time, but still in accordance
with the regulatory duty cycling. The lowest aggregated duty cycle of 0~\% means that the particular device turns off the transmissions completely.

\section{System Model} \label{sec:model}

Consider $M$ devices connected to a single LoRaWAN gateway. Each device is assigned a
spreading factor to use for transmission by a network server.  We account for the
interference through the collision model, where collision occurs when two or more devices
try to transmit simultaneously in the same channel using the same spreading factor. We
also consider a LoRa-only configuration, in this work, such that no interference from other technologies is present. Different spreading factors are considered to be
entirely orthogonal. A fixed payload
size is assumed. 
We further assume that all devices are class A and have successfully joined the network
and transmit the messages without acknowledgement so that there are no downlink
transmissions. Due to the absence of acknowledgements, retransmissions are not
considered.  

Among all sub-bands, a device is given a subset of the sub-bands. Enumerate these
sub-bands 1 through $c$. Let $n_i$, $i=1..c$ and $\delta_i$, $i=1..c$ be the number of
channels and the duty-cycle\footnote{$\delta_i$ is a normalized value between [0, 1].} in
sub-band $i$, respectively. 
As described in the
specifications \cite{loraspec20015} and in the source code of the reference
implementation of a LoRa/LoRaWAN device\footnote{https://github.com/Lora-net}, the
scheduling of a LoRaWAN transmission happens as follows:
\begin{enumerate}
\item A device waits until the end of any receive window.
\item A device waits for any off-period due to aggregated duty cycling.
\item A device checks for available sub-bands, i.e., ones that are not unavailable due
    to regulatory duty cycling:
    \begin{enumerate}
    \item A channel is selected uniformly randomly from the set of channels in all available sub-bands.
    \item If there is no free sub-band, the transmission is queued in the first free
        sub-band. A random channel in that sub-band will be selected.
    \end{enumerate}
\end{enumerate}

A transmission, limited by the duty cycle $\delta$, with a transmission period $T_\mathrm{tx}$ infers a holding period, which, including the transmission itself is given by:
\begin{align}\label{E:holdingtime}
T_\mathrm{hold} = T_\mathrm{tx} + T_\mathrm{tx}\left(\frac{1}{\delta}-1\right) = T_\mathrm{tx}\frac{1}{\delta}.
\end{align}

The \textit{service rate} is the inverse of the holding time,
$\mu=\delta/T_\mathrm{tx}$. Sub-bands can have different duty cycles and in turn different service rates.
Let $\lambda$ be the generation rate of packets for a device. When several sub-bands are defined for the device the sub-band for the next transmission is selected according to the step 3-a) and 3-b). We define \textit{service ratio} $r_i$ as the fraction of transmissions carried out in the $i-$th sub-band.


%
%
\section{Analytical Model} \label{sec:ana}

In this section the analytical models for latency and collision probability are presented. 

\subsection{Single Device Model: Latency} \label{S:analat}
The latency of a transmission is the time spent on processing, queueing, transmission of
symbols, and propagation. Assuming that the time for processing and propagation are negligible,  we have:
\begin{align}\label{E:totaldelay}
T_\mathrm{total} = T_\mathrm{tx} + T_\mathrm{w}.
\end{align}

We model the wait for reception windows and aggregated duty cycling ( steps 1) and 2) )
as a single traffic shaping $M/D/1$ queue. The service rate of this $M/D/1$ queue is the
slowest mean rate of service in step 1) and 2). For step 3), we model the regulatory duty cycling as an $M/D/c$ queue with heterogeneous servers, where each server corresponds to a sub-band. The waiting time $ T_\mathrm{w}$ for a transmission and the service ratio of each sub-band can then be found from queue theory.

The waiting time, $T_w$, due to regulatory duty-cycling can be calculated for asymmetric $M/D/c$ queue\footnote{The term ``asymmetric" captures the heterogeneous service rates of sub-bands.} that models step 3) of the scheduling procedure, 
but as it is easier to model and compute on a $M/M/c$ queue relative to a $M/D/c$ queue,
we use the rule of thumb that the waiting line of a symmetric $M/M/c$ queue is
approximately twice that of an $M/D/c$ queue \cite{newoldmdc} to simplify our analysis.
Our simulations show that this is a good approximation also for asymmetric queues.

The waiting time in sub-band $i$ is then:
\begin{align}\label{E:waitdelay}
T_{w_i} = \frac{p_{busy,all}}{(\sum^c_{i=1}\mu_i+\lambda)\cdot 2},
\end{align}
where $p_{busy,all}$ is the Erlang-C probability that all servers are busy. The
transmission latency in each band can then be found from Eq.~\eqref{E:totaldelay}.  The
mean latency is given as a weighted sum of the transmission latencies in each sub-band,
where the weights are given by the service rate of each sub-band.

The fraction of transmissions in sub-band $i$, $\lambda_i$, is the product of the holding-efficiency of the sub-band (fraction of time it is held) and the service rate of the band throughout that period. Then the service ratio is:
\begin{align}\label{E:serviceratio}
r_i = \frac{\mu_i}{\lambda} \cdot({1-p_{i\mathrm{,idle}}}),
\end{align}
where $p_{i\mathrm{,idle}}$ is the probability that the sub-band $i$ is idle. 

The service ratio can be expressed in short-hand forms for the two extreme cases of all sub-bands being available or busy all of the time.
When all sub-bands are available at the time of a transmission
the channel of transmission is selected uniformly
from the set of all channels as per step
3-a):
\begin{align} \label{E:rlimit1}
\lim_{\lambda\rightarrow 0}(r_i) = \dfrac{n_i}{\sum_{j=1}^cn_j}.
\end{align}

In the case that all sub-bands are unavailable at the time of a transmission
the transmission is carried out in the next available sub-band as per step 3-b):
\begin{align} \label{E:rlimit2}
\lim_{\lambda\rightarrow \mu_c}(r_i) = \dfrac{\delta_i}{\sum_{j=1}^c\delta_j}.
\end{align}

In order to describe $r_i$ between these extremes, we must find
$p_{i\mathrm{,idle}}$. Hence, we wish to find the steady-state probabilities given a
Markov model of the sub-band selection behaviour. For this purpose the model of a
\emph{jockeying}\footnote{Jockeying: A packet changes queue to a shorter queue if, upon
the end of service of another packet, it is located in a longer queue.} $M/M/c$ queue
from \cite{generalJockey} has been adopted. The Markov model of the jockeying queue has a
limited state space since, by definition, the difference in the number of queued
transmissions in any two sub-bands may not be larger than one. This allows us to put up a
matrix $\mathbf{A}$ containing all state transition probabilities, which can be used to
evaluate the steady state probabilities, $\mathbf{P}$, by solving the linear system
$\mathbf{A}\cdot\mathbf{P} = 0$. It also allows adoption of a Markov model for LoRaWAN
device behaviour, which is step 3) in the sub-band selection, by introducing state
transition probabilities based on the number of channels in each non-busy sub-band in
$\mathbf{A}$.

The jockeying queue does has a limited state space. As in \cite{generalJockey} we
approximate the model by making it finite by limiting the queue sizes to 1000. The model
now allows us to compute the steady state probabilities of all states; Amongst them
$p_{busy,all}$ and $p_{i\mathrm{,idle}}$. Then $T_{w_i}$ and $r_i$ can be calculated from
Eq.~\eqref{E:waitdelay} and Eq.~\eqref{E:serviceratio}.
%
Note that waiting times are lower for a jockeying queue than a regular queue. Hence
applying the rule of thump for approximation of an $M/D/c$ queue from a $M/M/c$ queue on
a $M/M/c:jockeying$ queue, will yield a \emph{lower} latency approximation of the $M/D/c$
queue.

\subsection{Multiple Devices Model: Collisions}

In this work we assume that no devices are making use of the optional acknowledgement
feature of LoRaWAN. Hence there is no DL in the model and as another consequence no
retransmissions occur upon collision.

It is empirically found in \cite{2017arXiv170404257M} that spreading
factors are not orthogonal in practice and, due to capture effect, one transmission may
be received successfully if the power of the wanted transmissions is sufficiently greater
than the interfering one. Unfortunately, at present there is no model of capture effect
in LoRaWAN and in this work, for simplicity, we assume that all channels and all SFs are
orthogonal. When two or more transmissions happen in the same channel, using the same
SF, at the same time, they collide.  This means we can model the access scheme as
multichannel ALOHA random access, as in \cite{mikhaylov2016analysis,
adelantado2016understanding,
augustin2016study}. 
Since there are 6 spreading factors defined for LoRaWAN, we have 6 sets of $n_i$ orthogonal Aloha-channels in sub-band $i$. The collision rate must be evaluated for each spreading factor.

We found
the service ratios of each sub-band in Section~\ref{S:analat}. Since the number of devices, the transmission time for the spreading factor being evaluated and the mean inter arrival time are known, we can calculate the load within a sub-band. The load within the
sub-band is spread uniformly over the channels allocated to that band.
Hence the traffic load of M devices, in sub-band i, given $SF_{i,j}$ is
\begin{align}
L(i,j) =\frac{\lambda\cdot r_i\cdot T_{\mathrm{tx,j}}\cdot M \cdot p_{\mathrm{SF}_{i,j}}}{n_i}
\end{align}
where $p_{\mathrm{SF}_{i,j}}$ is the percentage of all devices $M$, which use the $j$'th
spreading factor in sub-band $i$.

The collision probability is then
\begin{align}\label{E:colpr}
p_{\mathrm{col},i,j} = \exp\left({-2\cdot L(i,j)}\right).
\end{align}
In the paper, only unacknowledged UL transmissions are considered. So DL
limitations and retransmissions are not considered in this work. Therefore the outage is
caused by collisions can be quantified by our model.





%

\section{Performance Evaluation} \label{sec:eval}

In this section the latency given by Eq.~\eqref{E:totaldelay}, the service ratios given
by Eq.~\eqref{E:serviceratio} and the collision probability given by Eq.~\eqref{E:colpr}
are evaluated numerically. The evaluation is done for SF 12 based on 125~kHz channels,
50~bytes payload, 13~bytes overhead, code rate 4 and preamble length
$n_\mathrm{preamble}=8$.

\begin{figure}[tb]
\centering
\includegraphics[width=.95\columnwidth]{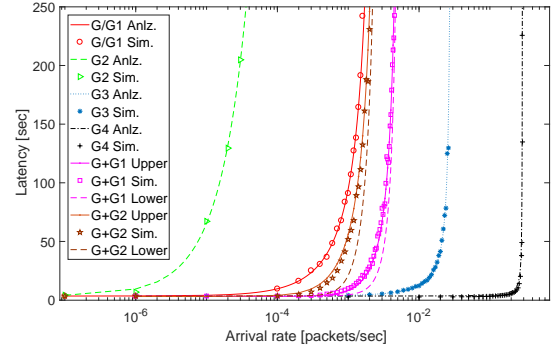}
\caption{Latencies on all sub-bands and combinations of sub-bands. Results denoted \emph{Upper} and \emph{Lower}  are calculated using ordinary $M/M/c$ model and jockeying model, respectively.}
\label{F:allSBlag}
\end{figure}

The latency including the transmission time and the waiting time due to regulatory duty
cycling as a function of arrival rate are depicted in Fig.~\ref{F:allSBlag}. The latency is plotted for stand-alone usage of each sub-band (G to G4) and for two
sub-band combinations (G+G1 and G+G2). 

The analytical approximation using Eq.~\eqref{E:waitdelay} for a heterogeneous $M/M/c$
queue provides a tight upper bound of the cases for the multiple sub-bands (G+G1 and
G+G2) and a tight approximation for the single band cases. The latency obtained by the
jockeying $M/M/c$ queue provides a lower approximation. The results show that lower latencies and higher capacities can be achieved for sub-bands with higher duty-cycles and combinations of bands with high duty-cycles.

\begin{figure}[tb]
\centering
\includegraphics[width=.95\columnwidth]{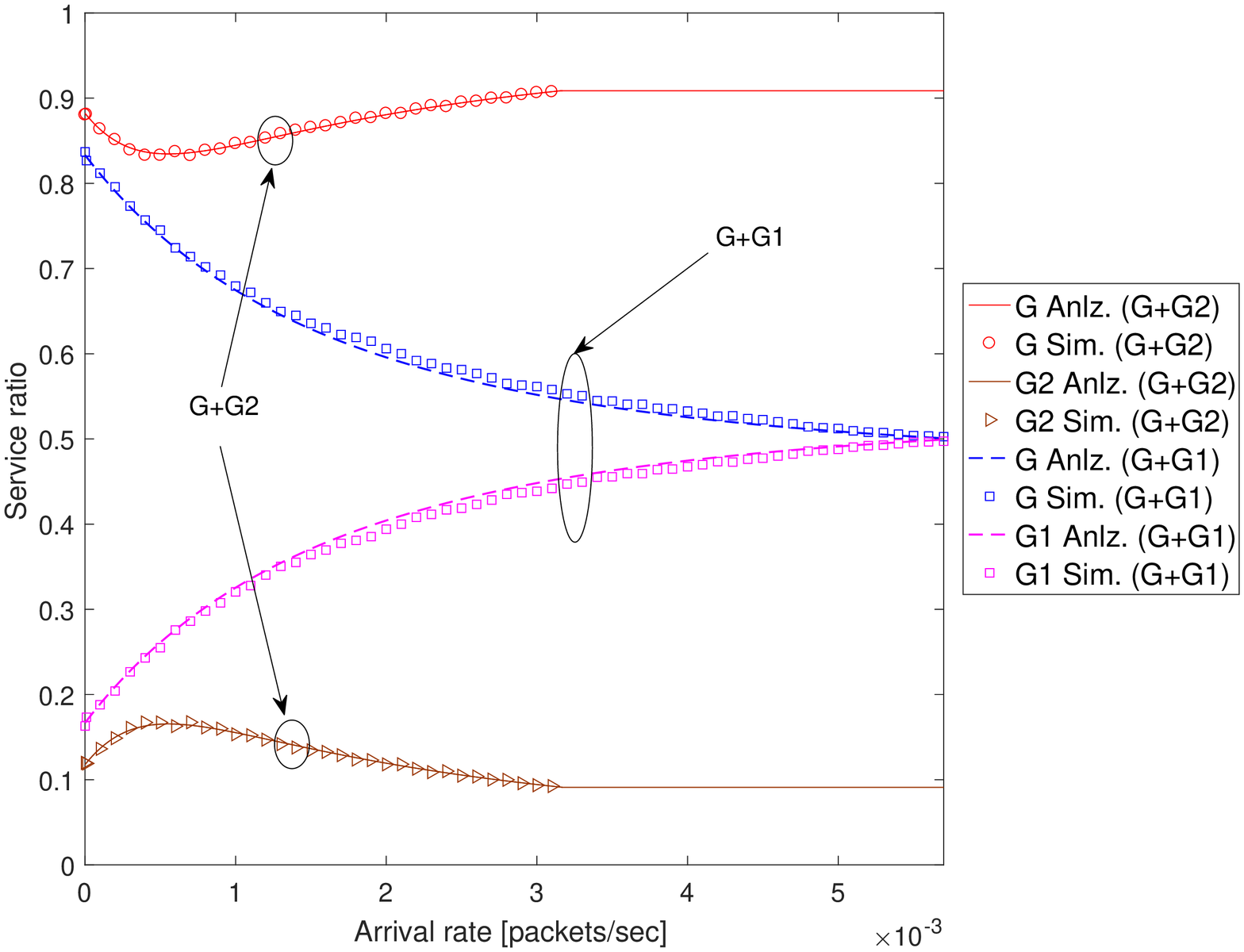}
\caption{Service ratios for G+G2 and G+G1.}
\label{F:SB1SB3col}
\end{figure}

The service ratios for the cases with combined sub-bands are plotted in
Fig.~\ref{F:SB1SB3col}. We see that combining G with G1 and G2, respectively, leads to
very different service ratios for the bands. G contains 15 channels and G1 contains just
3, but they have the same duty-cycle. The combination of G and G1 yields the service
ratio limit $15/(15+3) = .834$ for G for low arrival rates, but since the duty-cycling is
the same for the sub-bands we have the limit $.01/(.01+.01)$ for a high arrival rate.
%

\begin{figure}[tb]
\centering
\includegraphics[width=.95\columnwidth]{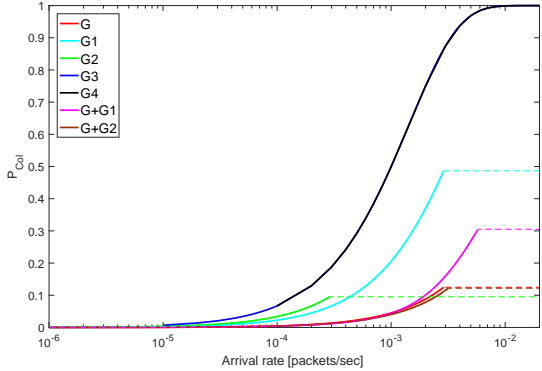}
\caption{Sub-band collision rates for 100 devices transmitting with SF$_{12}$.}
\label{F:allSBcol}
\end{figure}

The consequence of the sub-band pairing becomes evident by the collision rates depicted in Fig. \ref{F:allSBcol}. We see that the collision rate for G+G1 is larger than that of G alone or G+G2. This is due to the traffic not being spread equally on the channels for high arrival rates for G+G1. Since the limits of G+G2 are much closer, the load is spread more uniformly over the channels at high arrival rates and we see a drop in collision rate by adding the sub-band.
%
Notice that the
devices reach their capacities $\mu_c$ before the collision rate comes close to 1. In
this way duty-cycling limits the collision rate for each band, allowing for more devices
to share the band, but in practice arrivals beyond the capacity of each device would be
dropped.

From Fig.~\ref{F:allSBlag} it seems that the sub-band with the highest duty-cycle, G4, is attractive as it delivers low latency even at very high loads. However, when collisions are taken into account, we see that the sub-band has a very high collision rate as it only contains a single sub-channel. On the other hand, the lowest duty-cycle is found in sub-band G2, which has relatively high latency even at low loads, but with a lower collision rate than G4.

\begin{figure}[tb]
\centering
\includegraphics[width=.95\columnwidth]{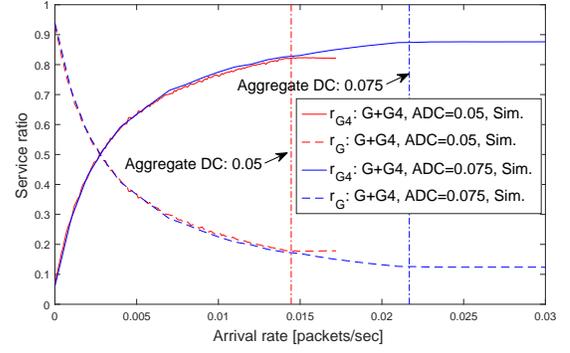}
\caption{Effect of aggregated duty cycle on service ratios.}
\label{F:aggDC}
\end{figure}



In Fig.~\ref{F:aggDC} the service ratios for G+G4 with an aggregate duty cycle of 0.05 (equivalent to a service rate capacity of the $M/D/1$ queue is 0.0146) and an aggregate duty cycle of 0.075 (equivalent to 0.0219) are plotted. The introduction of the aggregated duty cycle ($M/D/1$ queue) was found to effect the regulatory duty cycle queue ($M/D/c$ queue) by the service capacity, which freezes the sub-band service ratios of the regulatory queue and limits the obtainable latency.

%
%


\section{Concluding Remarks} \label{sec:conc}
A model for evaluating the performance of LoRaWAN UL in terms of latency and
collision probability was presented.
The numerical evaluation was done for EU868 ISM band regulations,
but the analysis is also valid for other bands utilizing duty cycling, such as the CN779-787 ISM band.

Short-hand forms for the limits of $r_i$ were presented. Equalizing the limits keeps the collision rate of sub-band combining at a minimum. The trade-off for this is a higher latency.
The traffic shaping effect of aggregated duty-cycling was shown and may be used as a built-in tool for collision-latency trade-off when combining sub-bands.

The UL model presented in this work, can be combined with DL models for Class A, B and C LoRaWAN devices and more sophisticated collision models to give insight into the bi-directional performance in LoRaWAN.





\bibliographystyle{IEEEtran}  
\bibliography{lora}

\end{document}